# Energy Efficient and Resilient Infrastructure for Fog Computing Health Monitoring Applications

**Ida Syafiza M. Isa, Mohamed O.I. Musa, Taisir E.H. El-Gorashi, and Jaafar M. H. Elmirghani**

*School of Electronic and Electrical Engineering, University of Leeds, Leeds, LS2 9JT, UK*

**ABSTRACT**

In this paper, we propose a resilient energy efficient and fog computing infrastructure for health monitoring applications. We design the infrastructure to be resilient against server failures under two scenarios; without geographical constraints and with geographical constraints. We consider a heart monitoring application where patients send their 30-seconds recording of Electrocardiogram (ECG) signal for processing, analysis and decision making at both primary and backup servers. A Mixed Integer Linear Programming (MILP) model is used to optimize the number and locations of the primary and backup processing servers so that the energy consumption of both the processing and networking equipment are minimized. The results show that considering geographical constraints yields a network energy consumption increase by up to 9% compared to without geographical constraint. The results also show that, increasing the number of processing servers that can be served at each candidate node can reduce the energy consumption of networking equipment besides reducing the rate of energy increase of networking equipment due to increasing level of demand.

**Keywords**: Energy Efficiency, ECG Signal, Fog Computing, Gigabit Passive Optical Networks (GPON), Health Monitoring, Mixed Integer Linear Programming (MILP), Primary Processing Server, Backup Processing Server, Resilience.

## 1. INTRODUCTION

The advancement of fog computing at the edge of the network has aided cloud computing and services to provide better quality of services (QoS) due to reduced latency. This is essential for health applications as low latency is a key requirement for delivering efficient services to patients. Furthermore, fog computing also helps in reducing the energy consumption in core and cloud networking infrastructures [1] − [16] under increasing applications' traffic. In our previous work in [17], we have shown that, using fog computing to run Electrocardiogram (ECG) monitoring applications within the time constraint imposed by American Heart Association (AHA) to save the heart patients, has contributed 68% total energy savings compared to the traditional cloud computing method. The consideration of fog computing to perform local processing at the edge network also improves services resilience. This has been studied in [18], where based on their simulation, fog computing improves network resilience by offering local processing at the network edge and provides better response time compared to a cloud-only architecture especially for interactive requests. Several research efforts used the concept of virtualization to provide network resilience; As in [19], the utilisation of backup servers has been improved by 40% by using virtualisation which allows the sharing of backup servers in the geo-distributed data centres. However, the shared protection scheme proposed in this work increases the latency of the secondary path between the primary and backup servers and requires high reserved bandwidth. Meanwhile, the work in [20] studies the benefit of relocation (i.e. location of primary and backup can differ) in terms of total cost of both servers and network capacity. The study reveals that with relocation, the cost of both servers and links capacity is reduced when considering protection against single link failures. Besides, the benefits of relocation are more pronounced for sparser topologies.

In this work, we propose a resilient infrastructure for health monitoring applications where the placement of backup servers together with the primary servers are optimized at the network edge so that the total energy consumption is minimized. The rest of the paper is organized as follows: Section 2 describes the proposed resilient infrastructure of fog computing to serve the ECG monitoring application. Section 3 presents and analyses the results and Section 4 gives the conclusions.

## 2. RESILIENT FOG COMPUTING INFRASTRUCTURE FOR HEALTH MONITORING APPLICATIONS

The architecture of the resilient fog computing infrastructure for healthcare applications over the GPON network consists of four layers as shown in Figure 1. The first layer is the IoT layer that comprises of IoT devices to monitor the health of the patients and to send data to the network. The second layer is the access layer which aggregates the health data from layer 1 using Long-Term Evolution for Machines (LTE-M) base stations. Each base station is connected to an Optical Network Terminal (ONT), and all the ONTs are connected to a single Optical Line Terminal (OLT). This layer is divided into several clusters where each cluster has one OLT connected to the ONTs of that cluster, and fog computing processing resources that can process and analyze health data, available only at the ONTs and OLTs. Although the Fog computing processing servers belong to a certain cluster, they are able to process the health data from any other cluster. Placing the processing servers (PSs) at the ONT can reduce the energy consumption of networking equipment as it is closer to the patients, but it will increase the required number of PSs. Meanwhile, utilizing the PSs at the OLTs reduces the number of required PSs as it is the shared point between the base stations in the cluster but will increase the energy consumption of networking equipment [17]. The third layer is the metro layer which aggregates and forwards data between the PSs in the access network, while the fourth layer is the core layer that is integrated with the central cloud and is used to permanently store the analysed health data.

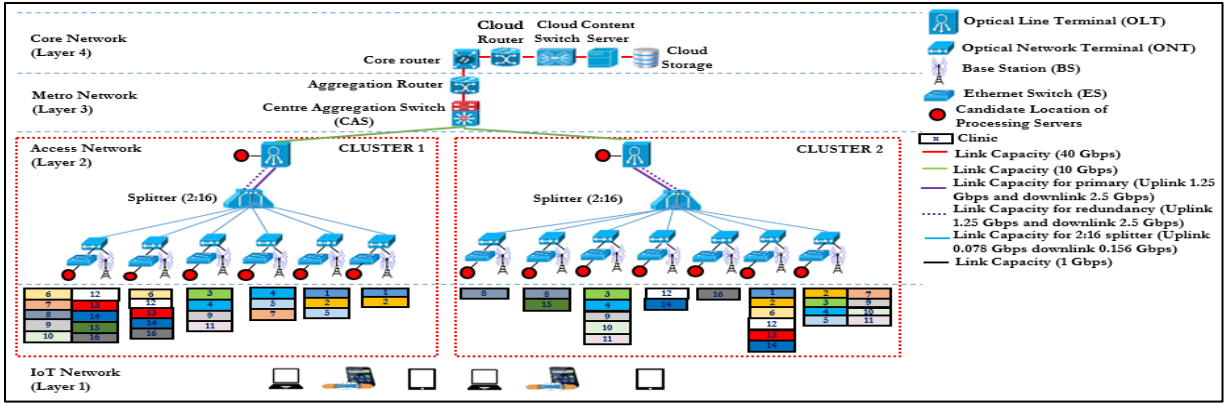

*Figure 1: The resilient fog computing infrastructure for health monitoring applications*

To improve the resilience of the health monitoring applications, we consider a 1+1 protection scheme where two servers, a primary server and a backup server are used to serve the health monitoring applications concurrently. We investigate two scenarios related to the geographic location of primary and backup servers offering two levels of resilience. The first scenario considers protection of servers without geographical constraint. In this scenario, the primary and backup processing servers can be placed at the same node. On the other hand, the second scenario considers protection of servers with geographical constraints where the primary and backup processing servers are not allowed to be placed at the same node. The latter offers higher levels of resilience compared to the former as node failures at a given location are not improbable.

## 3. PERFORMANCE EVALUATION

In this section we optimize serving the ECG monitoring application patients at both primary and backup processing servers in the fog architecture described above considering a scenario with 16 clinics that have a total of 300 patients and 13 Long Term Evaluation (LTE) base stations at the access layer using the locations at West Leeds as a case study. The 13 LTE base stations are selected based on the nearest distance between the available base stations (BSs) and the clinics. We consider a case study with two clusters and the clinics are connected to up to two nearest BSs in each cluster as shown in Figure 1. For example, clinic 13, shown in red, is connected to two base stations in cluster 1, and also a single base station in cluster 2.

In this work, 30 seconds ECG recording signal ($\Pi$) with a size of 252.8 kbits was retrieved from the MIT_BIT Arrhythmia database [21], [22] as it offers accurate results for the analysis, as recommended in [23]. Patients send their ECG signals to the network to be processed and analysed at both primary and backup servers at the fog layer. An experiment was conducted using MATLAB with parallel processing functions to determine the relationship between the processing time of the signals and the number of patients, utilizing the Pan Tompkins algorithm to perform the processing due to its high accuracy [24]. Based on the results, the duration of processing and analysis at a given number of patients ($Pat$) is: $\tau p = 0.002 \cdot Pat + 4.6857$, where maximum $Pat$ that can be served at a single server is considered to be 20% of the total number of patients from the 16 clinics. Based on the analysis results in the experiment, the size of the processed and analysed data was found to be 256 bits. This result will be sent from the primary and backup servers to the cloud for permanent storage, but only one copy will be stored. The same principle applies to the data that is sent to the clinic from both servers.

The energy consumption of networking equipment and processing is calculated based on the timing constraints set by the American Heart Association (AHA) [25]. As in our previous work [17], we consider $\tau t = 4$ minutes as the maximum duration that includes the time to record the 30-second ECG signal, $\tau m$, the transmission time to send the ECG signal from patient to both servers, $\tau max$, the processing and analysis time of the ECG signal, $\tau p$, and the feedback time to the clinics, $\tau b$. To determine the available time to transmit the ECG signal to the PS, $\tau max$, we calculate the time of both processing and analysis, $\tau p$ based on the maximum number of patients that can be served by a PS ($Pat$) and the feedback time to send the analysed ECG data to the clinics, $\tau b$ for $\tau t$ equal 4 minutes.

The feedback time is calculated as follows. We determine the maximum number of patient ($MaxP$) that can be served by the PSs at each candidate node. Due to the limited spaces at fog nodes, we limit the maximum number of PSs, $N$ that can be connected at each candidate node to 2 and 4 while each PSs can serve a maximum of $Pat$. Therefore, the $MaxP = N \cdot Pat$. Then we determine the minimum capacity between the candidate location of PSs at the access layer to the LTE BS (i.e. uplink between ONT and OLT). As the link capacity will be shared by the maximum number of patients the processing servers can serve at a node, we divide the link capacity by $MaxP$ to obtain the data rate for each patient to transmit the analysed data to the clinics ($\delta f$). However, in LTE BS, each user is given a minimum of 1 resource element (RE). Therefore, the minimum data rate that can be given to each patient is 336 bps per single RE when using Quadrature Phase Shift Keying (QPSK) modulation format. Due to this, we allocate the maximum number of RE ($RE_f$) to each patient while ensuring that the given data rate does not exceed $\delta f$, hence the data rate for feedback is $\delta b = \delta r \cdot RE_f$ while the feedback time is $\tau b = \alpha/\delta b$. In addition, we allocate maximum number of REs to each patient to send their ECG signal to the PSs while ensuring that the given data rate, $\delta a$ is higher than $\Pi/\tau max$. Due to this, the transmission time to send the ECG signal to the processing servers for processing and analysis is $\tau a = \Pi/\delta a$. Note that, the link capacities are shared by multiple applications. Therefore, for healthcare applications, we consider 0.3% of the maximum available link capacity as in [17]. The data rate for storage task ($\delta c$) is calculated by dividing the minimum uplink capacities between the candidate location of PSs and the cloud storage (i.e. link between the ONT and OLT) by $MaxP$. Hence the transmission time to send the analysed ECG data to cloud for permanent storage is $\tau c = \alpha/\delta c$.

We consider five scenarios with 20%, 40%, 60%, 80% and 100% of the total number of patients in the 16 clinics to investigate the impact of increasing number of patients on the energy consumption of networking equipment and processing. This was done while considering the two protection scenarios and different number of allowed PSs at each candidate node.

As in our previous work [17], the power profile considered for all networking equipment and processing servers consists of idle power and a linear proportional power. The maximum power and idle power of the LTE base station are 528 Watt and 333 Watts working at 0.3 Gbps [26], respectively, while for Ethernet switches are 3.52 Watt and 0.57 Watts working at 16 Gbps [27], respectively. The maximum power, idle power and capacity of other networking devices are as in [17]. The processing servers selected in fog have Intel Core i5-4460 with 3.2 GHz CPU with maximum power and idle power of 180 Watt and 78 Watts [28], respectively. Note that, the networking devices are shared by multiple applications, thus only 0.3% of the idle power is considered while the PS and Ethernet switch are dedicated for the healthcare applications.

The number of processing servers and their locations given the aforementioned constraints are optimized using a MILP model, where the objective is to minimize the consumed energy in all networking and processing equipment in the used architecture. The MILP was run using IBM CPLEX software suite running on a high performance computing platform with 24 cores processor and 256 GB of RAM.

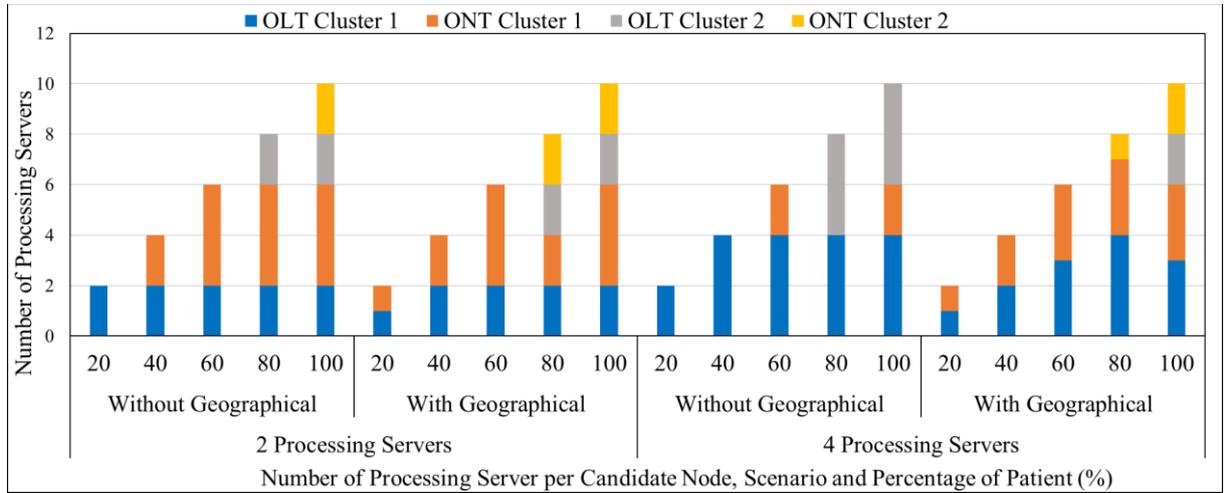

*Figure 2: Processing server placement in the network*

Figure 2 shows the MILP results for the number and optimal locations of primary and backup servers, comparing the first scenario without geographical constraints with the more resilient second scenario considering geographic constraints. The results are shown when 2 servers are available at each candidate node, and when this number is doubled to offer more processing capacity per node.

The results show that the OLT is always chosen to place the processing servers as it is the nearest shared point to the patients (the OLT is connected to all BSs of the same cluster) which reduces the number of required PSs and the number of hops to transmit the ECG signal to the PSs. The results also show that, the PSs are placed at only one cluster when the percentage of patients considered in the network is equal to or less than 60% as all patients can be served by the BSs in one cluster only. Therefore, to reduce the number of utilized networking equipment in the network, the ONT is selected to place the remaining PSs which cannot be allocated in the OLT at the same cluster. However, increasing the percentage of patient to 80% and 100% has resulted in utilizing BSs, ONTs, and OLTs from both clusters.

Figure 2 also shows that, when considering geographical constraints, the number of nodes utilized to place the PSs is higher than the scenario without geographical constraints. This is because with geographical constraints, the primary and backup servers are not allowed to be placed at the same location. However, this depends on the number of patients and number of PSs that can be connected at each candidate node. Besides, increasing the number of PSs that are available at each candidate node can also reduce the number of nodes to place the PSs for both scenarios. For server protection with geographical constraints, at a demand level of 80% of patients, and 4 PSs available at each candidate node, the OLT and ONTs of cluster 1 are occupied first, and the remaining demand is sent to the ONT of cluster 2. This is to reduce the total amount of data traversing the network as ONTs are directly connected to the patients. However, when the demand level increases to 100%, the OLT and the ONT of both clusters are used.

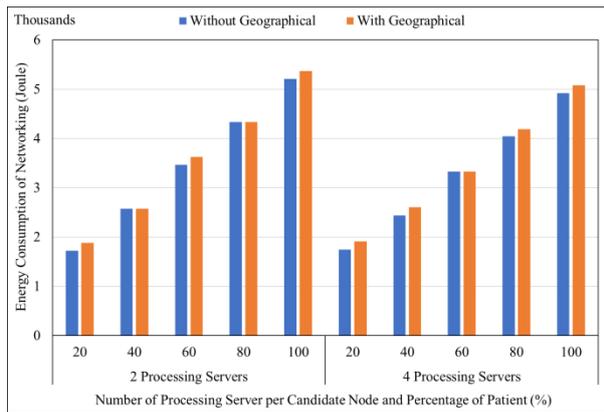

*Figure 3. Energy consumption of network*

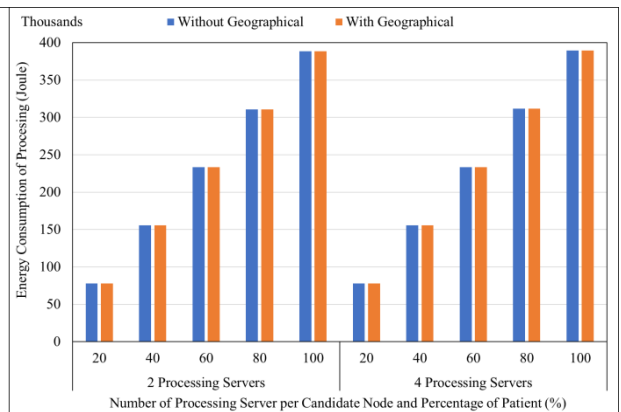

*Figure 4. Energy consumption of processing*

Figure 3 and 4 show the energy consumption for networking equipment and processing, respectively for an increasing level of demand when the number of PSs allowed at each candidate node is 2 and 4 in each case. The energy consumption increases by 9% when opting for the geographical constraint, at demand level of 20%, and 2 available processing servers. This is because, when the geographical constraint is considered, more nodes are utilized to place the PSs, hence this will increase the total number of utilized networking equipment. This increase in energy is the penalty for having a higher level of resilience. However, this penalty can be reduced by increasing the number of processing servers at each candidate node. Doubling the number of processing servers also reduces the rate of energy consumption of networking equipment as the demand level increases. This is because, more PSs can be placed at the same node, hence reducing the number of utilized networking equipment. However, at demand level of 40% and 80% the energy consumption of networking has not increased. This is a positive result as the same level of resilience does not incur any energy penalty. This is because at this specific demand level, the same number of nodes are used to place the PSs in both scenarios.

Figure 4 shows that the energy consumption of processing increases as the percentage of patients increases for both scenarios when 2 PSs and 4 PSs can be served at each candidate node as increasing the number of patients increases the number of PSs proportionally. For both number of processing servers per candidate node, the energy consumed is equal for the both resilience levels as the same number of servers will be utilized regardless of their location.

## 4. CONCLUSIONS

In this paper, we proposed a resilient energy efficient Fog based processing architecture for health monitoring applications. We optimized the placement of both primary and backup processing servers to process and analyze ECG signals from patients at the network edge under two scenarios; with and without geographical constraints. The results show that considering geographical constraints can increase the energy consumption of networking equipment compared to the case without geographical constraints as a penalty for increasing the system resilience. However, increasing the number of processing servers that can be served at each candidate node can reduce the energy consumption of networking equipment and reduce the increase in networking equipment energy consumption due to the increase in number of patients. Meanwhile, the increasing energy consumption of processing is only affected by the increasing number of patients in the network as more processing servers are utilized in the network.

**ACKNOWLEDGEMENT**
The authors would like to acknowledge funding from the Engineering and Physical Sciences Research Council (EPSRC), through INTERNET (EP/H040536/1), STAR (EP/K016873/1) and TOWS (EP/S016570/1) projects. The first author would like to thank the Ministry of High Education of Malaysia and Universiti Teknikal Malaysia Melaka (UTeM) for funding her PhD scholarship. All data are provided in full in the results section of this paper.